# Cost Projections for High Temperature Superconductors

Paul M. Grant and Thomas P. Sheahen
EPRI, Palo Alto, CA 94304 and SAIC, Gaithersburg, MD 20878

*Abstract* --- **It is generally argued that for high-temperature superconductors (HTS) to be cost-competitive in power applications, the wire will have to sell for about \$10 per kiloampere×meter (\$10/kA×m) for operation at 77 K (e.g., NbTi costs around \$1/kA×m and $Nb_3Sn$ around \$8, each at 4.2 K). Given what is already known about the critical current performance of Pb-stabilized Bi-2223 (BSCCO), this cost target may be extremely difficult to realistically achieve for silver-sheathed BSCCO produced by the oxide-powder-in-tube (OPIT) technique. In this paper, we examine the cost of component materials, add reasonable estimates for labor and related costs, and arrive at a likely cost/performance (C/P) figure. We also estimate the capital cost of a factory to produce HTS conductor by a particular coated conductor method, and calculate the necessary production-output and performance parameters necessary to manufacture 10 km/yr of wire and its associated C/P. Our results indicate that the real C/P seen by the customer will remain substantially above this \$10/kA×m target for some time to come.**

## I. Introduction

For over a decade, the emphasis of applied research on high-temperature superconductivity (HTS) has been placed on improving the properties of the materials, especially the critical current $J_C$. Until recently [1], little attention has been given to considerations of manufacturing cost, because other concerns and obstacles have always been of much more immediate concern. There seems to be a basic presumption that if researchers can find a way to do something at all, then others will find a way to drive down manufacturing costs and make a profit from it.

For power applications [2] of HTS, where many amperes are to be transported over sometimes meters, sometimes miles, there exists competition with traditional low-temperature superconductors (LTS) such as NbTi and $Nb_3Sn$, because the latter can perform the exact same functions whenever the temperature can be reduced into the liquid helium range (4 K). The very high cost of such refrigeration has kept many superconducting power applications off the market for years, despite their demonstrated technical feasibility [3]. The attitude of utilities and power equipment manufacturers has been one of reluctance to accept superconducting technology because of the prohibitive costs (from their point of view) in light of performance gained.

For electrical wire, the figure-of-merit for comparing costs of different materials at a particular operating point has been dollars per kiloampere×meter (\$/kA×m). This reflects the twin purposes of a wire, namely, to carry high currents over great distances. Notice that it is current, not current density, that matters here; from a practical point of view, any conductor carries an "overhead" burden of sheathing, insulation, or (in the case of conductors coated onto a nonconducting substrate) the thickness of extra material. For aluminum and copper wire, the distinction is a few percent; for superconductors, the distinction is at least a factor of two, and often far greater, reaching up to 100 or more in adverse cases. The usual critical current density, $J_C$, must thus be set aside in favor of the "engineering" critical current, $J_E$, where all these burdens are included. The \$/kA×m number is what the user has to design to and pay for whatever kind of conductor that takes electricity from one place to another.

In addition to the capital cost of buying wire, the user has to spend money for installation, maintenance, repair, and the continuing cost of cooling the wire to its operating temperature. Any reduction in the refrigeration expense due to running at a higher temperature is of value to the customer, and this can result in an increase of the acceptable capital (manufacturing) cost of HTS wire.

Given what is known today about the comparative tradeoffs between refrigeration and wire manufacturing costs, and recognizing that $Nb_3Sn$ costs \$8/kA×m, and NbTi \$1/kA×m, it is generally thought throughout the HTS community [4] that it will be necessary for HTS conductor to sell for about \$10/kA×m in order to have substantial penetration of power application markets. Of course, the exact price sensitivity varies from one application to another, and there are wide error brackets, but the general feeling is that anything greater than \$50/kA×m will satisfy only niche requirements.

The remainder of this paper is devoted to presenting an analysis of how well the cost/performance (C/P) target of \$10/kA×m is likely to be met by BSCCO (Pb-stabilized $Bi_2Sr_2Ca_2Cu_3O_{10}$, or Bi-2223) wire tapes made by the OPIT (oxide-powder-in-tube) method, and by YBCO ($YBa_2Cu_3O_7$, or Y-123) coated conductors. We anticipate many of our conclusions will be controversial and perhaps some even in error. This is to be expected given that much critical data is presently being held close by the various companies and institutes involved for quite understandable propriety and competitive reasons. Nevertheless, we believe it is now time to commence discussion of cost/performance challenges that confront both process developers, manufacturers and end users of high temperature superconducting wire.

## II. BSCCO WIRE

To put the HTS target number in perspective, we employ the "Sokolowski Plot" shown in Fig. 1 [5]. This conveniently displays operating current vs. cost for various combinations of material, temperature and magnetic field. The plot is also divided into regions of constant C/P in \$/kA×m as indicated by the various diagonal lines. Note only $Nb_3Sn$ and NbTi lie to the right of the \$10/kA×m line, and that several of the reported values are manufacturer's targets, not achieved results. Presently, HTS tapes available on the "open market" are quoted between \$800-1200/kA×m.

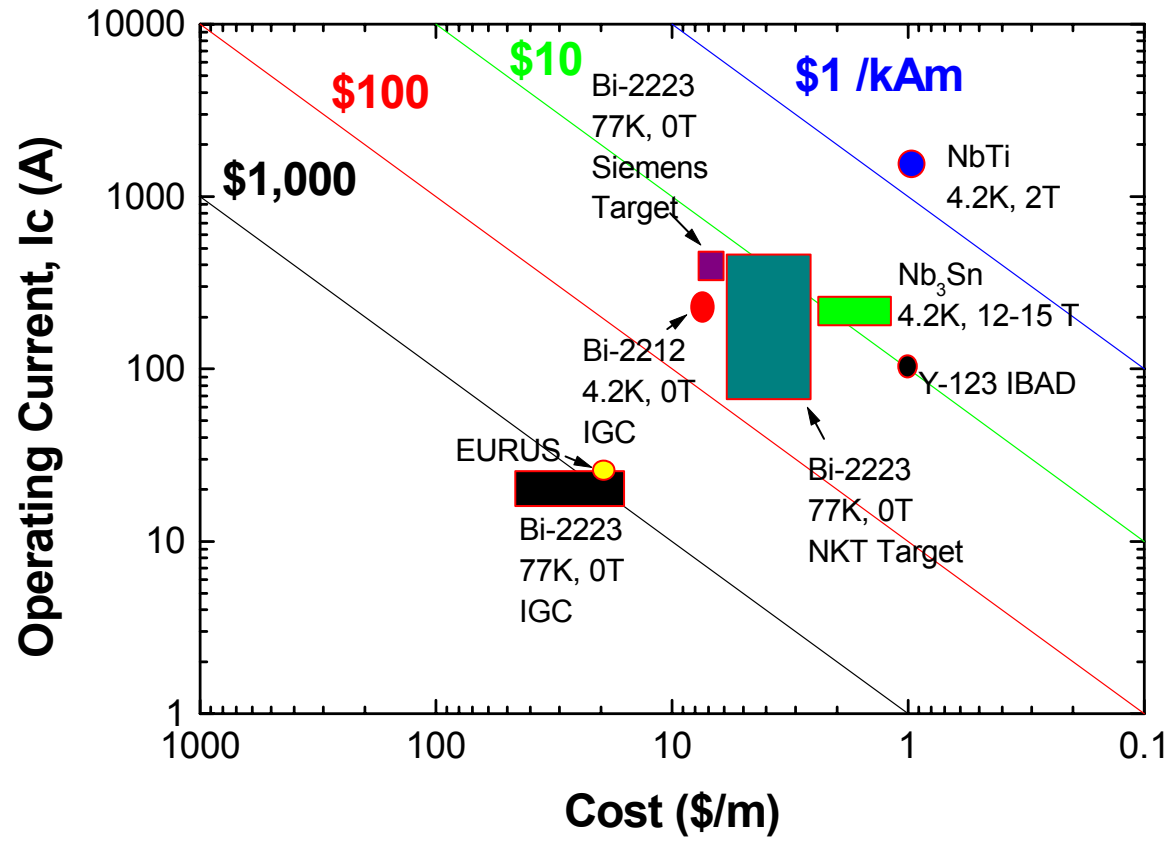

Fig. 1. The "Sokolowski Plot" of a number of HTS and LTS wire embodiments. Data were obtained courtesy of the manufacturers as labelled in the plot. When comparing performance, note the different operating points, i.e., temperature and field. The diagonal lines represent demarcation of various constant C/P values in \$/kA×m.

As mentioned at the Applied Superconductivity Conference, ASC-96, two years ago [6], the cost of silver sheathing for BSCCO carries a very severe price penalty. The calculation is straightforward, and might readily be left as an exercise for the reader; but nonetheless we will present our version now.

Assume a typical powder-in-tube configuration has a silver tube 2 mm in diameter, with an inside diameter of 1.15 mm filled with BSCCO. The ratio of cross-sectional areas is such that silver is 2/3 and BSCCO 1/3; so the *volume fraction of superconductor* is $\lambda = 1/3$. As the tube is stretched and thinned, that ratio doesn't change. When a bundle of tubes is combined to form a multifilament wire and crushed into a tape, still the ratio doesn't change.

The silver tubes could perhaps have a thinner wall, but the need for thermal protection [6] against burnout during quench argues for having more than half the wire made of non-superconducting material [7]. It would be dangerous to run above $\lambda = 0.5$. In most previous applications, such as NbTi for accelerator magnets, $\lambda = 0.4$ is common. For any wire with adequate thermal protection, the silver cost component will certainly be nontrivial. Figure 2 gives an idea of the variation in silver futures for the early part of 1998.

For numerical simplicity, let's say the HTS material in a given BSCCO tape has a critical current density $J_C = 2 \times 10^4$ A/cm$^2$ [8]. Thus 200 amperes can be carried over one mm$^2$. The silver surrounding that much BSCCO is roughly 2 mm$^2$, and a segment one meter long therefore contains 2 cm$^3$, which weighs 21 grams. At an average of \$5/troy-ounce, one meter of silver sheath costs \$3.38 yielding a C/P for 200 A of \$16.88/kA×m. Depending on volume ordered or internally produced, the cost of the superconductor material runs between \$0.34-1.37/cm$^3$ at stoichiometric density for Bi-2223 [9]. We then need to add to silver a C/P for the HTS material on average of \$4.28/kA×m for a rounded-down total of \$21/kA×m. This leaves the C/P over 100% above target *for materials alone*, with no indirect manufacturing costs (labor, capital, etc.) yet included.

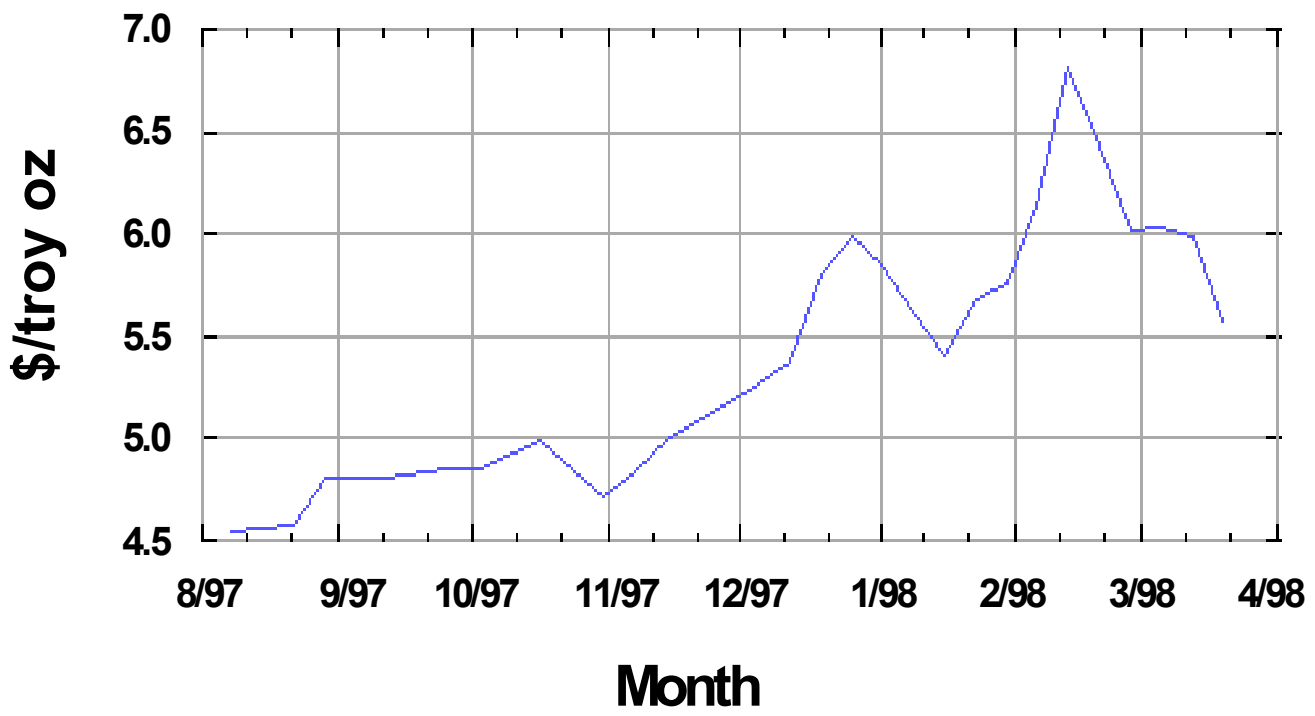

Fig. 2. Variation in May 1998 silver futures since August 1997. The peak in February displays the "Warren Buffet Effect" (the large volume silver purchases that month by Berkshire Hathaway) the and is suggestive of the volatility of silver prices against moves by large investment groups.

What are the future prospects for silver supply and prices [10]? In 1997, total worldwide demand for silver was 27 million kilograms, outstripping supply by 6 million kg. Demand for silver has exceeded supply since 1990, not exactly a good sign for a new technology that will make even greater demand on reserves, in the ground, recycled and stored as bullion. One of the largest consumers of silver is the photography industry (7.2 million kg/yr of which 1.4 million kg/yr is recycled), and it is sometimes remarked that the replacement of silver halide-based photographic technology by digital methods will lower silver prices to the benefit of other uses, e.g., power applications of superconductivity. Yet in 1997, photographic uses of silver increased 3%, a trend which is expected by industry analysts to continue for some time for amateur and some professional applications.

The annual production of NbTi by *IGC* is about 20,000 km/yr [11]. Annual installation of underground transmission cable (three phase circuits) is around 3×60 km/yr in the US. Assuming a form factor of 20 to acount for layering and helicity, this would require 3,600 km/yr of HTS tape. Let's say the total potential annual market for practical HTS tape is twice this total number (i.e., complete takeover of all commercial applications of LTS and new HTS opportunities), or nearly 50,000 km/yr in the US. For BSCCO/OPIT tape of the form factor we have been assuming (3mm$^2$ and $\lambda = 1/3$), this scenario would consume about all the silver currently recycled by the photographic industry, a situation likely to adversely affect silver prices, assuming other demand remains constant (perhaps the wisest investment strategy for a utility is to insert Ag-based HTS wire technology as rapidly as possible, subsequent to salvage at a huge profit 40 years from now after the discovery of room temperature superconductors!).

Returning now to additional factors affecting BSCCO wire C/P, estimates for labor and overhead (L&O) range from $1-5/m [9], which transform into $5-25/kA×m in our $J_C$ = 20,000 A/cm$^2$ example above. Given this very wide bracket (right now, the production lines are manned mainly by PhDs!), let's split the difference in L&O which now raises the overall C/P for OPIT BSCCO to $37/kA×m [12].

Next, some real bad news. There are further considerations which must be taken into account when arriving at a practical number for an "end use" C/P, which we will call "derating factors." These factors have to be inserted because it is the *operating current*, not the laboratory $I_C$, which, from the viewpoint of the end user, must comprise the normalization unit in C/P. There are at least four elements which contribute to lowering the real current that the wire can carry, and two more, one related to strength and ac loss and the other to marketing, which can raise the cost. All are multiplicative. First of all, the “voltage drop criterion” by which $J_C$ is defined is commonly taken to be 1 μV/cm for HTS wires, but in real applications, it will be necessary to stay about 20% (depending on the particular value of n in $E \sim J^n$, presently about 10-15) below this "conventional $J_C$" in order to lower the voltage drop below 0.1 μV/cm, the criterion for establishing $I_C$ in LTS wire, otherwise resistive losses become untenable. Second, It is a wise precaution to assume that most applications will take place in a magnetic field of at least 0.1 tesla, which reduces $J_C$ by 15% (80% if the field is aligned perpendicular to the ab plane). Third, for manufactured lengths over 1 km, a factor of about 3 may be lost, at least as indicated by publicly available data (*American Superconductor* [13] maintains they can achieve only a 5% derating over long lengths, but to the knowledge of the authors, supporting data has not been published). Fourth, it is generally assumed, especially in the transmission/distibution cable community, operation under ac conditions will lower the current capacity by a factor of as much as 2.5. Fifth, we must remember that metallurgically pure silver (not even sterling!) is not used in the actual BSCCO/OPIT manufacturing process for reasons of strength and reduction of ac loss. This additional alloying and/or processing does not come for free. We speculate such "finishing touches" will add perhaps 50% or more to the "commodity price" of silver. Finally, wire manufacturers presumably want to make some profit -- 30% over basic manufacturing cost seems fair to us at this time.

These various derating factors, and their products, are summarized in Table I.

TABLE I
VARIOUS C/P DERATING FACTORS

| Derating Factor | This Paper | *ASC* [13] |
|---|---|---|
| 1.0 → 0.1 μV/cm | 1.2 | |
| 0.1 T Magnetic Field | 1.8 | 1.15 (B ∥ ab plane) |
| Length ≥ 1 km | 3.0 | 1.05 |
| ac Operation | 2.5 | |
| Ag Treatment | 1.5 | |
| Profit | 1.3 | |
| Cumulative Derating | 31.6 | 7.1 |

The right hand column in Table I represents informal communication from *American Superconductor* on what are felt by them to be more appropriate numbers for field and length derating. The rows left empty do not necessarily represent their agreement or disagreement with our estimates in the middle column.

Thus the cumulative derating applied to our previous figure of $36/kA×m now raises the true C/P to somewhere between approximately $230-1040/kA×m. Admittedly, there is room for manuever here. For example, not all applications involve ac, and a considerable reduction in C/P may be in reserve, enough perhaps to offset the cost of ancilliary equipment for dc/ac conversion for some utility transmission /distribution system applications. In addition, a factor of about two improvement can be accomplished simply by reducing the operating temperature to around 66 K (77 K was assumed the cryogenic operating point throughout our discussion so far). Nonetheless, the bottom line is that the greatest opportunity for reducing C/P lies in significantly increasing the fundamental $J_C$. What are the prospects this can be done?

Some idea can be obtained through a study of Fig. 3 which shows the results of magnetoptically imaging the flux penetration of an externally applied magnetic to a BSCCO/OPIT filament followed by an inversion of the Biot-Savart integral equation to obtain the actual critical current paths and distributions therein.

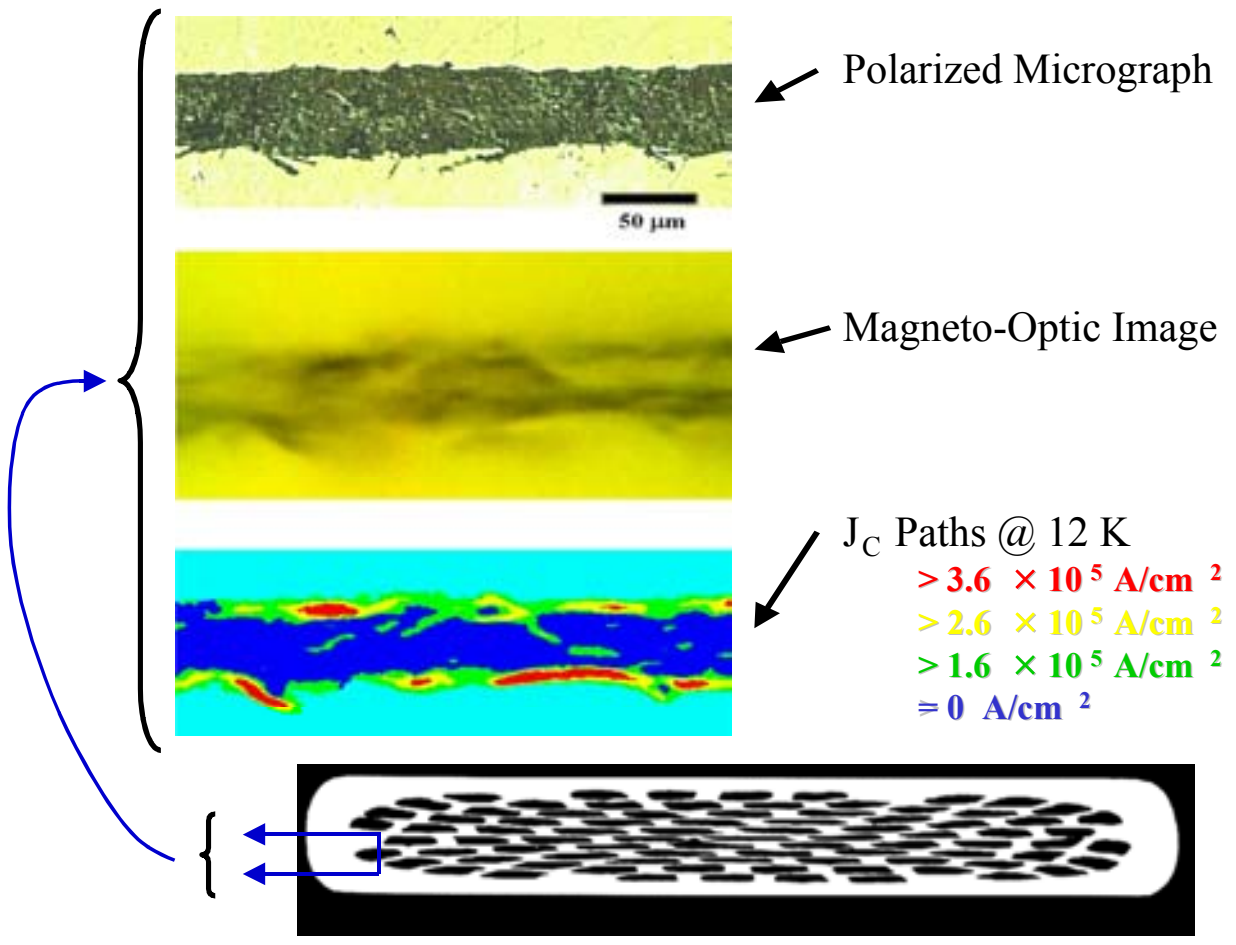

Fig. 3. Cross-sectional microscopic and magnetoptical (polar Kerr effect) images (MOI) of BSCCO filaments in OPIT tape. The lowest picture shows

the lateral cross-section of a typical multifilamentary (~80 here) BSCCO/OPIT tape such as manufactured by *IGC* and *ASC*. The upper three stacked images are of a longitudinal "take out" of one such filament, the topmost being a polarized micrograph (note 50 µm scale bar) clearly showing the granular nature of the filament. Next is the MOI of the flux penetration of a small external magnetic field applied perpendicular to this same area cooled to 12 K, followed by a map of the numerical inversion of the Biot-Savart equation yielding critical current paths. Multifilamentary tape photo courtesy of *ASC* and the rest of the figure is taken from the U. Wisconsin Applied Superconductivity Center home page [14].

The truly startling consequence of Fig. 3 is that the overwhelming bulk of the filament (shown in blue) carries *zero critical current!* Most of the critical current is preferentially transported near the Ag/BSSCO interface, a result well-known in the field [2, 15]. In this particular sample, the contiuous current path exceeded 160,000 A/cm$^2$ (green and yellow), with a few small regions (red) reaching 360,000 A/cm$^2$.

Can we use the example of Fig. 3 to estimate an "as good as it gets" outcome for BSCCO/OPIT wire? The data were taken at 12 K, but a reasonable way to extrapolate to 77 K might be as follows: 1) the relative $J_C$ at all temperatures and fields is distributed more or less as shown in Fig. 3, i.e., with the maximum at the Ag/HTS interface; 2) the maximum $J_C$(77 K, 0 T) for epitaxial BSCCO films is roughly one million A/cm$^2$; 3) one might assume it is possible to achieve perhaps 1/3 this magnitude near the Ag surface in BSCCO tape; and 4) let's just assume that would be 360,000 A/cm$^2$, the same value seen in Fig. 3 for 12 K (after all, the data of Fig. 3 is now three years old, and recently $J_C$ values near this figure have been observed by at least one institution [16]). The "as good as it gets" scenario for BSCCO/OPIT would then be played out if this $3.6 \times 10^5$ A/cm$^2$ could be realized throughout the entire cross-section of the filament.

How long might it take before such could be accomplished? To date, $J_C$ in meter-scale lengths has increased linearly in time since the birth of BSCCO wire technology around 1991…a kind of linear "Moore's Law" which in fact has been christened "Malozemoff's Law" after the *ASC* scientist who first observed this trend. The slope of $J_C$ vs. t is about 9200 A/cm$^2$/yr, and this allows, under an assumption Malozemoff's Law will continue to hold, an estimation of when "as good as it gets" for BSCCO is realized. This speculative scenario is exhibited in Fig. 4.

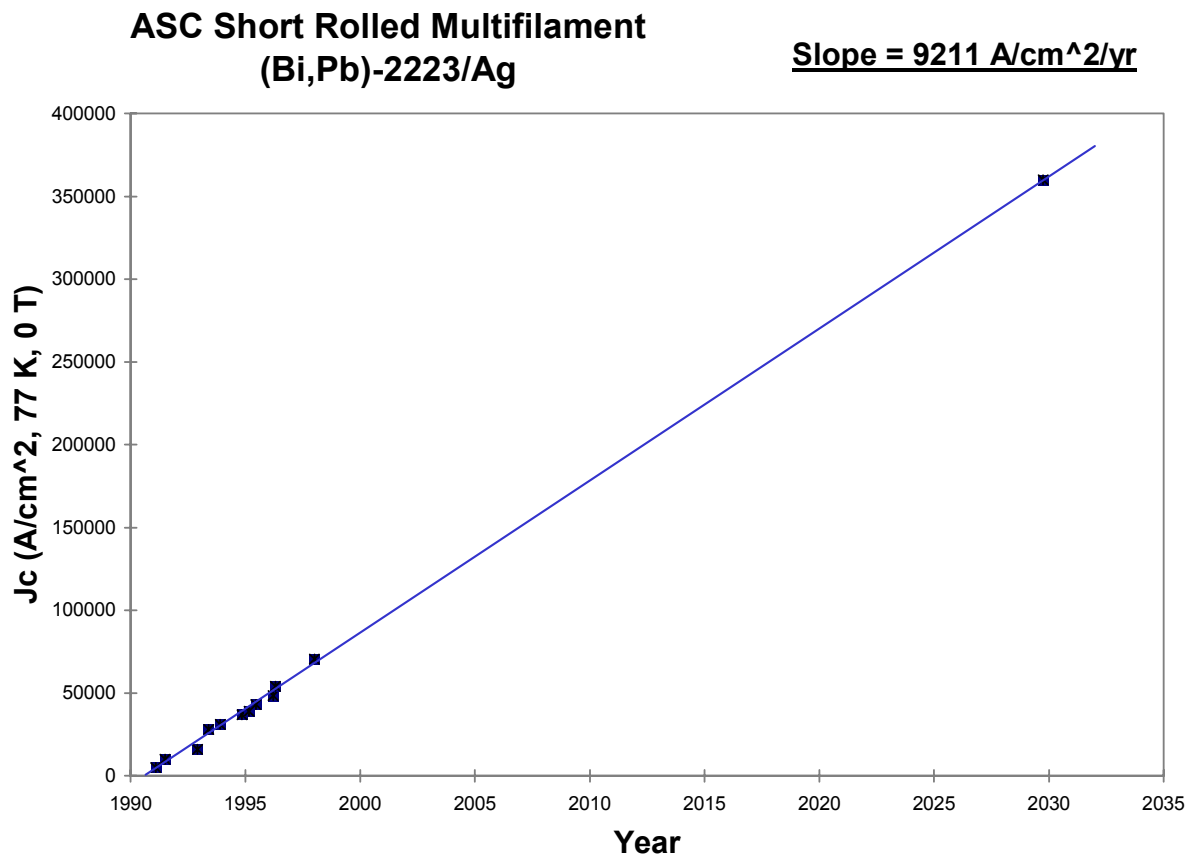

Fig. 4. "Malozemoff's Law" showing the linear increase in $J_C$ between 1991 and the present. Assuming its continuing trend, maximum $J_C$ of 360,000 A/cm$^2$ for BSCCO/OPIT will arrive near 2030.

We see that BSCCO/OPIT wire, subject to our assumptions and the continuance of Malozemoff's Law, will top out around the year 2030…quite a long time to wait. We suspect $J_C$ vs t will deviate from linearity long before then. Which way, of course, will prove crucial to its future application.

Some estimation of expected trends for the "real" C/P can be made by combining our derating exercise above with the extrapolation of Malozemoff's Law just discussed and displayed in Fig. 4. Figure 5 shows the result.

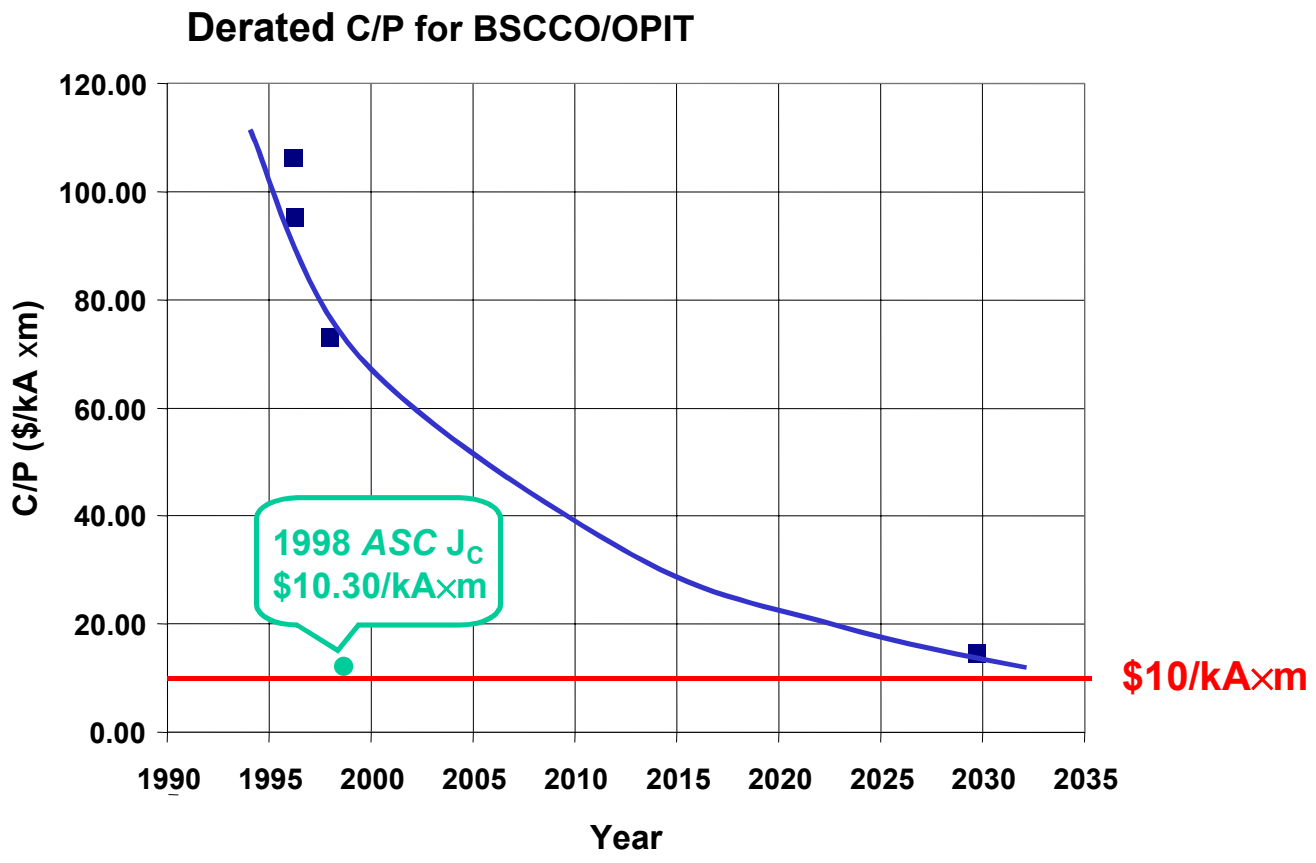

Fig. 5. Prediction of the decrease in BSCCO C/P per year employing the data of Fig. 4 since about mid-1995, the geometrical, Ag-filling, materials, and L&O numbers of our earlier example, all derated by a factor of 7.1. The "call-out" displays the "unburdened" C/P associated with the record value of $J_C$ = 70,500 A/cm$^2$ announced by *ASC* in 1998 [17].

If we view \$10/kA×m as merely a "laboratory benchmark," one could plausibly claim it has already been reached (see the balloon call-out in Fig. 5 pointing to the 1998 70,500 A/cm$^2$ record announced by *ASC* earlier this year [17]). We see that achievement of the generally accepted target price of \$10/kA×m as a practical market objective will remain out of reach for a very long time, perhaps many decades into the future. Even if the ac loss burden (2.5) is removed by some future large demand for dc cable, the \$10 figure is unlikely be reached before 2020…and remember…up to now, our analyses have not even included capital-cost-of-plant amortization. Therefore, it appears that the real cost of BSCCO wire will remain far above the "target price" of \$10/kA×m for a very long time, certainly many decades into the future.

## III. YBCO Coated Conductors

The hope for this "second generation" of high temperature superconductor wire relies on thin films of high $J_C$ performance YBCO deposited on inexpensive substrates. A very thin veneer (a few microns) of silver may be needed on top of the YBCO, but the thickness (and therefore cost) would be much less than for BSCCO. The main advantage to YBCO coated conductors is that the $J_C$ values in the YBCO film are hoped to be ultimately around $10^6$ A/cm$^2$, a full order of magnitude above the current BSCCO range.

As there is not presently *any* production of YBCO coated conductors ongoing, not even in pilot line mode, it is thus necessary to imagine a future manufacturing plant, using a particular process, and then estimate the costs associated with materials, labor, capital equipment, etc., in order to arrive at the total expense of buiding, running and maintaining the plant. For any given total output, it is then an easy calculation to arrive at the \$/kA×m cost of this conductor. Alternately, since this is an imaginary plant to begin with, it is possible to hold the cost fixed at \$10/kA×m, and then ask how much the total annual output must be -- in other words, how fast must the production lines run to achieve break-even.

The latter approach has been taken in a report by Chapman [18, 19], who looked at two potential YBCO manufacturing process in considerable detail. Envisioning a hypothetical factory where YBCO is vapor-deposited onto a nickel substrate following the ORNL RABiTS™ process, the various components of the production cost were carefully estimated. In that report, both pulsed laser deposition (PLD) and electron-beam (e-beam) methods were evaluated. Chapman found that PLD turned out to be far too expensive, but the e-beam process might be viable.

To achieve the \$10/kA×m target cost, it was necessary to produce 18,000,000 meters per year of coated conductor. This is comparable the typical production, mentioned earlier, of *Intermagnetics General Corporation* of 20,000 km/yr of NbTi superconducting wire in their large scale production facility in Waterbury, CT [11]. It is worth pointing out that this fabrication rate for NbTi by *IGC* is the result of nearly 30 years of diligent engineering and manufacturing R&D.

It will be instructive to examine the Chapman Report scenarios in more detail, because they provide valuable insight into the enormity of the scale-up problem that must be solved if YBCO coated conductors are to become commercially viable. The 18,000,000 meter plant envisions 30 parallel 1 cm tapes of nickel continuously moving across a 1/2 meter distance inside a chamber, all the while in a vapor of atomic Y, Ba, and Cu. In addition, the chamber must operate in a partial pressure of a few Torr of $O_2$, unless oxygenation is done in a subsequent process step.

The e-beam approach evaporates elemental atoms from targets within in the chamber. To gain necessary production capacity per year, Chapman assumes the rate of barium deposition in this hypothetical factory will be about 1 gram/sec, with copper and yttrium rates yielding an additional net mass transfer of 2 grams/sec. This is *comparable to painting a wall*, and, incidentally, does not include material deposited on the sides of the vacuum chamber and elsewhere. A typical e-beam target would be exhausted within a minute, and even very massive targets weighing several kilograms would need to be renewed roughly every hour, thus a continuous feed of stock material through vacuum-tight thrust bushings to the hearths would be required (this may not be as implausible as it sounds, as some methods of actinide isotope separation and enrichment for nuclear weapons and power reactor fuel are actually carried out this way). Both sides of the nickel tape are to receive a net deposition of 2 microns thickness during the 1/2 meter traverse of the chamber. That corresponds to a deposition rate of 1168 Angstroms/second. Low grain-angle boundary growth upon the substrate must be maintained for that full thickness.

As mentioned previously, Chapman's approach has been to fix the target price and scale the factory throughput to meet it.

If instead one reverses this thinking and imagines an e-beam based factory capitalized at perhaps \$33 million, producing a more modest throughput of 10 km/year, it is possible to then derive a \$/kA×m figure as follows: Using an *opportunity cost of capital* [20] of 18% (typical today), the plant costs \$6,000,000 annually even if *nothing* goes out the door. Adding reasonable estimates for labor and other operating costs (on a *per-year* basis, not on a *per-meter* basis as in section 2 above), as well as materials, brings the annual expense up to \$7,500,000. With an output of 10 km/yr, we then have an average cost of \$750/meter. If the conductor carries perhaps 400 Amps, again a typical number for coated conductors of the geometry commonly discussed, we arrive at \$1,875 /kA×m.

Each of these numbers can be massaged somewhat: lower capital cost, higher labor cost, more precise materials estimates, etc. No combination of changes can erase the two-order-of magnitude discrepancy between this and a factory producing "competitive" conductor. The key to improving price performance has to be to increase throughput of the factory. This implies speeding up the deposition process from what is feasible today at least tenfold, *and* throughput speed tenfold as well. That means developing a continuous process that gives uniformity and consistency of product to a degree as yet unapproached for high temperature superconductors.

In our opinion, it is difficult to conceive how either PLD or e-beam manufacturing methods will be able to realize a customer C/P in the range of \$10/kA×m. On the other hand, several groups have proprietary programs underway that use "wet and dry" chemical coating techniques, which, if success-ful, could dramatically lower production costs and quite possibly approach this number. As always, time will tell.

## IV. Summary

We believe we have made a strong case in our paper for thoughtful reconsideration of \$10/kA×m as a target market-entry cost/performance criterion for high temperature superconductivity wires and tapes. Indeed, if the only result is for us to have provoked controversy and discussion of this vital issue in HTS power applications, then our purpose has been served. It just doesn't look possible to achieve this number as a practical C/P result for a very long time to come, if ever…practical meaning manufacture and sale at a profit for a wide variety of power uses. We suspect, in fact we are convinced, there is no single C/P market-entry value whose realization would constitute a declaration of victory. Competition with LTS wires and devices will remain for a substantial period, especially, in our view, in very large applications such as high energy physics, and large generators and motors where helium cryogenics is both mature and improving.

Nevertheless, certain applications definitely benefit from a higher temperature refrigeration system, cables being the most dramatic example. The authors are aware of several instances where installation of low voltage, equivalent power distribution cables could enable a given utility to release urban real estate occupied by intermediate voltage step-down substations. The enormous savings and cash return therefrom could justify a C/P of perhaps as high as \$1000/kA×m (although this would surely be a niche business!).

We urge our colleagues in manufacturing companies to seriously consider issuing wire "specification sheets" so those of us in the end-user community can begin to intelligently engineer and financially plan our respective potential applications. This is most difficult if all we are given are artificial targets which have meaning only under laboratory conditions. Let's get going. There's a lot of work and, as always the case with applied superconductivity, a long road ahead of us.

## Acknowledgement

The authors have drawn on a large reserve of colleagues and network associations for the material used in this paper, and it would be impossible to cite them all. It would be also inappropriate to expose them to perceived support of our conclusions. Nonetheless, we are grateful to all, especially those willing to let us have a "peek under the covers" at otherwise proprietary issues.